\documentclass[12pt]{article}
\usepackage{epsfig,amstex,amssymb}
\begin{document}

\noindent \textit{Mat. Res. Soc. Symp. Proc. \textbf{651}: T5.3.1-T5.3.6 (2001)}\\

\noindent \textbf{\Large Mechanical, adhesive and thermodynamic \\
properties of hollow nanoparticles}\\

\noindent \textbf{U. S. Schwarz$^{1}$, S. A. Safran and S. Komura$^{2}$} \\
Department of Materials and Interfaces, Weizmann Institute, Rehovot 76100, Israel\\
${}^{1}$Max-Planck-Institute of Colloids and Interfaces, 14424 Potsdam, Germany\\
${}^{2}$Department of Chemistry, Tokyo Metropolitan University, Tokyo 192-0397, Japan

\section*{Abstract}

When sheets of layered material like C, WS$_2$ or BN are restricted to
finite sizes, they generally form single- and multi-walled hollow
nanoparticles in order to avoid dangling bonds.  Using continuum
approaches to model elastic deformation and van der Waals interactions
of spherical nanoparticles, we predict the variation of mechanical
stability, adhesive properties and phase behavior with radius $R$ and
thickness $h$. We find that mechanical stability is limited by forces in
the nN range and pressures in the GPa range. Adhesion energies scale
linearly with $R$, but depend only weakly on $h$.  Deformation due to
van der Waals adhesion occurs for single-walled particles for radii of
few nm, but is quickly suppressed for increasing thickness.  As $R$
is increased, the gas-liquid coexistence disappears from the phase
diagram for particle radii in the range of 1-3 nm (depending on wall
thickness) since the interaction range decreases like $1/R$.

\section*{Introduction}

\begin{figure}[tb]
\begin{center}
\begin{tabular}{cc}
\epsfig{figure=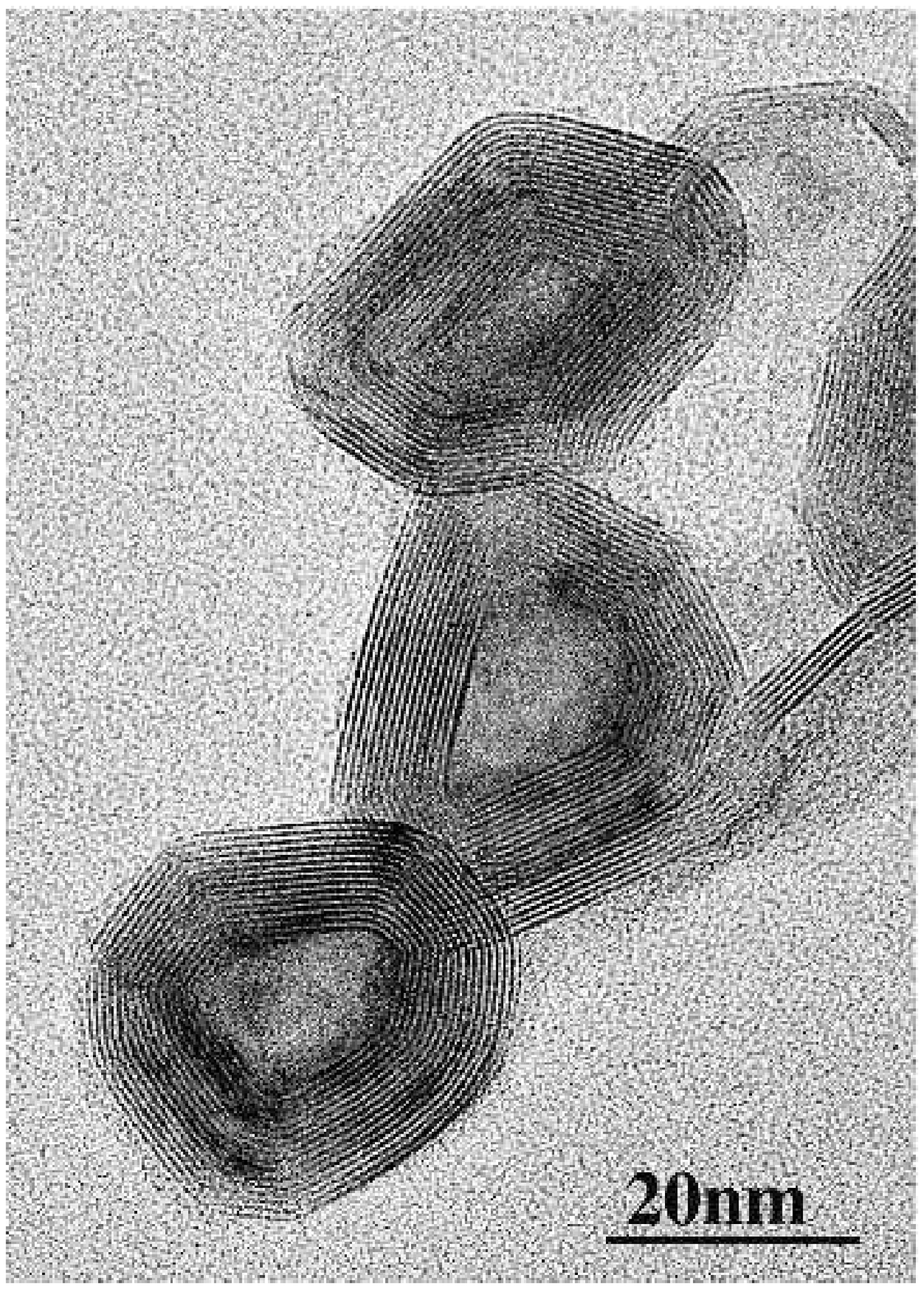,width=0.5\textwidth} &
\epsfig{figure=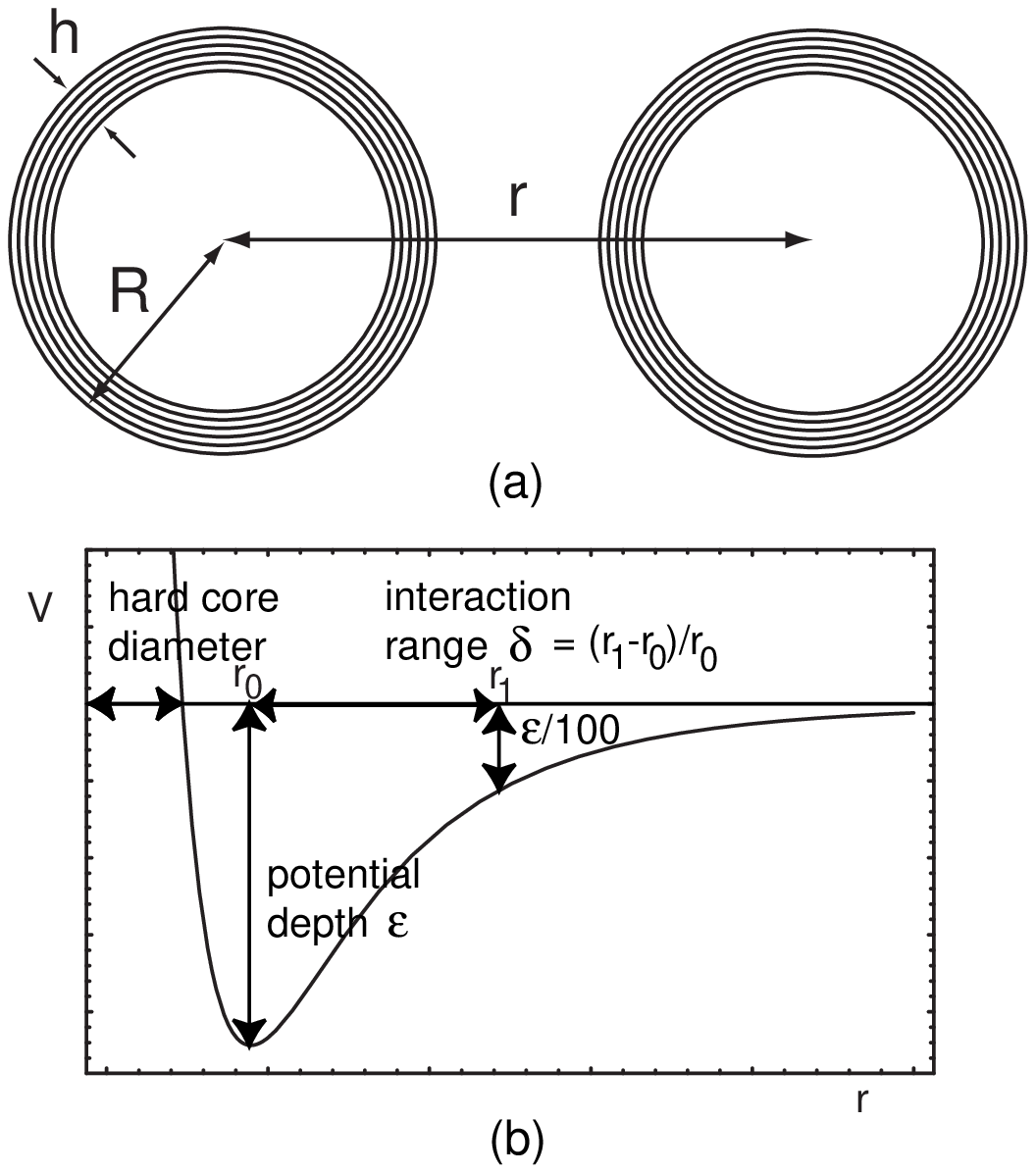,width=0.5\textwidth}
\end{tabular}
\end{center}
\caption{Left: Transmission electron micrographs of hollow MoS$_2$-nanoparticles
(courtesy of R. Tenne). Van der Waals interactions leads to strong
adhesion. The faceted shape is due to grain boundaries.  Right: (a)
Model for two spherical nanoparticles of radius $R$ and thickness $h$
each, interacting over a distance $r$.  (b) The effective interaction
potential $V$ as a function of distance $r$ is characterized by the hard
core diameter $2 R$, potential depth $\epsilon \sim R$ and interaction
range $\delta \sim 1/R$.}
\label{tem}
\end{figure}

The prototypical hollow nanoparticle is the buckyball C$_{60}$, which
crystallizes into \textit{fullerite}.  Many other morphologies for
carbon sheets have been found since the early 1990s, including
fullerenes C$_n$ of varying size, multi-walled fullerenes
(\textit{carbon onions}), and single- and multi-walled carbon
nanotubes.  Closed structures of carbon are formed in order to avoid
dangling bonds at the edges of finite sized carbon sheets. This
mechanism is generic for anisotropic layered material of finite size,
and up to now more than 30 other materials have been prepared as
hollow nanoparticles (see the reviews \cite{n:tenn95} and references
therein).  This includes the metal dichalcogenides MeX$_2$ (Me = W,
Se, X = S, Se), BN, and GaN. Usually \textit{inorganic
fullerenes} like WS$_2$ and MoS$_2$ are multi-walled.  In
Fig.~\ref{tem} we show transmission electron micrographs of
MoS$_2$-particles which were synthesized by solid-gas reaction.
Control of size and shape is rather difficult, as evidenced by the
irregularly faceted shapes. However, the methods used for the
synthesis of hollow nanoparticles are developing very rapidly, and it
is to be expected that control of size, thickness and shape will
become much better in the future.

Hollow nanoparticles combine covalent in-plane strength with flexible
out-of-plane bending of thin films, which results in high mechanical
stability.  They interact by van der Waals (vdW) forces, which for
example determine the material properties of fullerite \cite{n:giri92}
and the phase behavior of buckyballs \cite{n:cacc97}. VdW interactions
become even more important for larger fullerenes, carbon onions and
inorganic fullerenes; these materials, therefore, feature a large
degree of non-specific adhesion.  In particular, vdW adhesion can lead
to deformation of the hollow nanoparticles. For example, it has been
found that carbon nanotubes adhering to each other or to flat surfaces
show observable deformations for radii as small as 1 nm
\cite{n:ruof93}. In the presence of local surface features
like crystal steps, adhering nanoparticles will have spacially varying
electronic properties due to spacially varying deformation.

Hollow nanoparticles may lead to new applications in nanoelectronics
and -optics. In contrast to filled nanoparticles, they can also be
used for storage and delivery systems. They show superior mechanical
properties, like high flexibility, high tensile strength and light
weight, which will lead to applications as ultra-strong fibers.
Larger fibers can be produced using the cohesiveness provided by vdW
interactions.  Inorganic fullerenes have also been shown to offer
favorable tribological properties \cite{n:rapo97}. Detailed
investigations with the Surface Force Apparatus demonstrated that low
friction and wear is caused by material transfer onto the sliding
surfaces \cite{n:gola99}.

For these applications, a good theoretical understanding of the
physical properties of hollow nanoparticles is important.  Here we
discuss structural properties of spherical hollow nanoparticles in a
continuum approach that allows us to focus on unusual generic
properties which result mainly from geometrical effects. In
particular, we address mechanical properties, van der Waals
interaction, thermodynamic behavior and deformation by adhesion.  In
this paper, we focus on the main results of our work; more details can
be found elsewhere \cite{uss:schw00d,uss:schw00e}.

\section*{Mechanical properties}
  
During recent years, mechanical properties of fullerenes and carbon
nanotubes have been extensively studied at the level of molecular
calculations, that is by first principles, tight binding and force
field techniques. The advantage of molecular models is that they can
provide detailed quantitative predictions.  For larger systems like
multi-walled nanoparticles, their implementation becomes difficult due
to computer time requirements.  Here we use classical continuum
elasticity theory, which is asymptotically correct for large systems
and allows us to treat all these systems using the same framework. The
large advantage of this approach is that it provides insight into the
generic properties of these systems. Molecular calculations for carbon
nanotubes have shown that the predictions of continuum elasticity
theory persist well into the limit of radii smaller than 1 nm
\cite{n:robe92}. Recently this observation could be
verified also for BN and MoS$_2$ nanotubes \cite{n:hern98}.

The layered material discussed here usually has a hexagonal lattice
structure, for which elasticity theory \cite{b:land70} predicts that there
are only two in-plane elastic constants and the sheet is elastically
isotropic. Its deformation energy consists of stretching and bending
terms. For single-walled carbon sheets, the in-plane stretching
modulus $G \approx 3.6 \times 10^5$ erg/cm$^2$ and the out-of-plane
bending rigidity $\kappa \approx 3 \times 10^{-12}$ erg.  For both
MoS$_2$ and WS$_2$, $G$ is smaller by a factor $4$ and $\kappa$ is
larger by a factor $5$.  For multi-walled nanoparticles, the elastic
constants scale with thickness $h$ as $G
\sim C_{11} h$ and $\kappa \sim C_{11} h^3$, where $C_{11}$ is
the largest in-plane elastic constant of the corresponding layered
material \cite{n:srol94}. The values for $C_{11}$ are $1060$, $238$ and $150
\times 10^{10}$ erg/cm$^3$ for C, MoS$_2$ and WS$_2$, respectively.

Stretching can be avoided completely by bending the sheet into a
cylindrical nanotube. However, this is not possible for spherical
nanoparticles, which we discuss here. The same holds true for
saddle-like structures, and in general for all surfaces with non-zero
Gaussian curvature. On a microscopic level, the requirement for
non-zero Gaussian curvature also means that a flat sheet cannot be
bent into a sphere without introducing defects.  Continuum elasticity
approaches to fullerenes therefore explicitly treated the 12
pentagonal defects necessary to close a sheet of carbon hexagons
\cite{n:ters92}.  In order to avoid this complication, here we assume
that the curvature generating defects are distributed in a homogeneous
manner and result in a vanishing internal strain (like for a ping-pong
ball).  In this case, the hollow nanospheres can be modeled as elastic
shells with thickness $h$ and prefered radius $R$.

Mechanical collapse in tribological applications is likely to be
caused by either direct forces or high pressure.  In the latter case,
an inward buckling instability occurs, since the pressure energy
scales with a higher power of indentation $H$ than the restoring
elastic energy ($H^2$ versus $H^{3/2}$) \cite{uss:schw00d}. Mechanical
collapse is expected when indentation $H$ becomes of the order of
radius $R$. The critical pressure for collapse can then be estimated
to be $p_c \sim G^{1/4} \kappa^{3/4} / R^{5/2}$.  For buckyballs and
typical inorganic fullerenes, this yields values of 15 and 2 GPa,
respectively. For nanoparticles, the corresponding forces are in the
nN-range.

\section*{Van der Waals interaction and phase behavior}

Since vdW interactions are very difficult to treat on a molecular
level, a continuum approach is even more useful in this case.  It
is well known from colloid science that vdW interactions strongly
depend on the geometry of the system under investigation.  
Geometrical aspects of vdW interactions are well
treated in the Hamaker approach, in which an atomic Lennard-Jones
interaction is summed in a pairwise manner over the relevant geometry
and the molecular details of the interaction are lumped into the
effective Lennard-Jones parameters $\sigma$ and $\epsilon$.  For
carbon, $\sigma = 3.5 \AA$ and $\epsilon = 4.6 \times 10^{-15}$ erg
\cite{n:giri92}. The corresponding Hamaker constant $A$ 
is in the order of $10^{-12}$ erg. For two buckyballs, the Hamaker
approach for thin sheets leads to the \textit{Girifalco potential}
\cite{n:giri92}, which is considerably shorter ranged than a
Lennard-Jones potential.  Since a small range of attraction leads to
the disappearance of the gas-liquid coexistence from the phase diagram
\cite{n:gast83}, it has been suggested that buckyballs constitute the
first non-colloidal system for which the gas-liquid critical point
does not appear in the phase diagram.  However, large scale
simulations have shown that buckyballs indeed do show a gas-liquid
coexistence, which extends roughly from 1900 to 2000$^{\circ}$C
\cite{n:cacc97}.

For hollow nanoparticles in general, the effect of a finite thickness
$h$ becomes important. However, for nanoparticles in a crystal or
adhering to a flat substrate or to each other, the distance of closest
approach $D$ between interacting particles is on an atomic
scale. Under such conditions, the vdW interaction saturates on the
atomic scale $D$ with increasing thickness $h$. For example, the
surface energy for the interaction with a halfspace scales as
$\epsilon h / D^3$ and $\epsilon / D^2$ for thin films and halfspaces,
respectively, with the crossover occuring for $h = D$. Since $h
\gtrsim D$ in our case, the surface energy is $\epsilon / D^2$
independently of $h$ and no significant scaling with thickness is
expected.

\begin{figure}[tb]
\begin{center}
\epsfig{figure=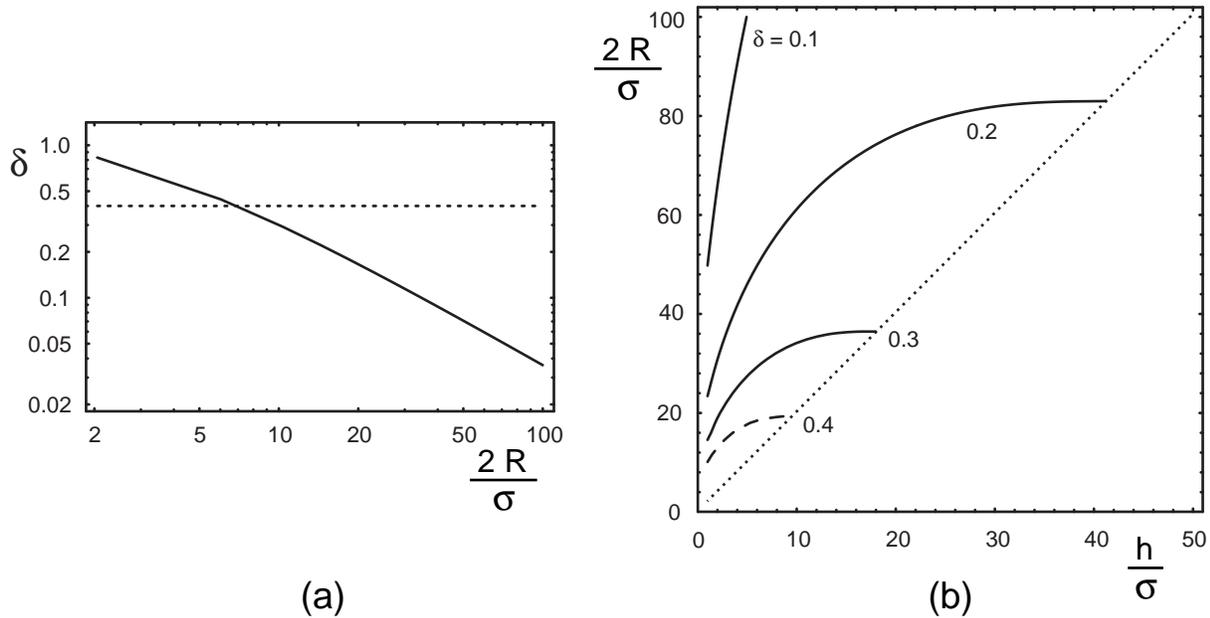} 
\end{center}
\caption{Numerical results for the interaction range $\delta$ 
(a) for single-walled nanoparticles as a function of radius $R$ and
(b) for multi-walled nanoparticles as a function of radius $R$ and
thickness $h$. $\sigma = 3.5\ \AA$ is the atomic length of the
Lennard-Jones potential. The gas-liquid coexistence is expected to
disappear from the phase diagram for $\delta < 0.4$ (dashed lines).}
\label{InteractionRange}
\end{figure}

For a more detailed investigation, one has to consider the general
case of two nanospheres of finite thickness interacting in a pairwise
manner by a Lennard-Jones potential \cite{uss:schw00e}. The
interaction energy can be written as $V = V_{R,R} + V_{R-h,R-h} - 2
V_{R,R-h}$, where $V_{R_1,R_2}$ is the interaction energy between two
filled balls of radii $R_1$ and $R_2$, respectively, which can be
calculated analytically for a Lennard-Jones potential. In order to
predict phase behavior, three quantities characterizing the resulting
potential are especially important: the hard core diameter $2 R$ sets
the density scale, the potential depth sets the temperature scale, and
the interaction range determines if a gas-liquid coexistence occurs
(compare Fig.~\ref{tem}).  Our full analysis has shown that for hollow
nanoparticles of radius $R$ and thickness $h$, the potential depth and
interaction range scale linearly and inversely with $R$. Both
quantities depend only weakly on $h$. In
Fig.~\ref{InteractionRange} we show our numerical results for the
interaction range as a function of $R$ and $h$. Recent work for the
Double-Yukawa potential suggests that the gas-liquid coexistence
disappears for an interaction range smaller than $\delta = 0.4$
\cite{n:teje94}. Our analysis predicts that this happens for $R = 12
\AA$ and $R = 35 \AA$ for single- and multi-walled nanoparticles,
respectively. For larger $R$, the critical point will disappear
from the phase diagram.

\section*{Deformation due to adhesion}

We consider the case where a spherical hollow nanoparticle is indented
a distance $H$ due to adhesion to a flat substrate. The vdW energy
gained on adhesion can be estimated to be $E_{vdW} \sim (A / D^2)
D H$ where $A$ is the Hamaker constant and $D$ an atomic cutoff
\cite{comment}.  The energy of deformation can be estimated to be 
$E_{el} \sim G^{1/2} \kappa^{1/2} H^2 / R$.  With increasing $R$,
deformation becomes more likely, since both the vdW energy and
deformability increase. However, deformability decreases rapidly with
increasing thickness $h$, since $G^{1/2} \kappa^{1/2} \sim
h^2$. Solving for the indentation $H$ and using the values given
above, it is found that for hollow carbon nanoparticles with few
walls, $H$ can be in the nm-range \cite{uss:schw00d}. However, for
typical inorganic fullerenes deformations are suppressed by the larger
shell thickness $h$.

Finally we briefly discuss a nanotube adhering to a flat substrate.
As explained above, nanotubes can be bent without being stretched.  On
deformation, one has to pay a bending energy per length of $E_{el} / L \sim
\kappa H^2 / R^3$.  The gain in vdW energy scales as $E_{vdW} / L \sim
A  H / D^{3/2} R^{1/2}$.  It follows that for carbon nanotubes
with small thickness $h$, the indentation $H$ can be in the nm-range
for even smaller values of radius $R$ than for hollow carbon
nanoparticles. 

\section*{Acknowledgments}

It is a pleasure to thank R. Tenne for many stimulating
discussions. USS was supported by the Minerva Foundation and SK by the
Ministry of Education, Science and Culture of Japan.


\begin{thebibliography}{10}

\bibitem{n:tenn95}
R.~{Tenne}, {\em Adv. Mater.}, \textbf{7}, 965 (1995) and
{\em Progress in Inorg. Chem}, \textbf{50}, 269 (2001).

\bibitem{n:giri92}
L.~A. Girifalco, {\em J. Phys. Chem.}, \textbf{96}, 858 (1992).

\bibitem{n:cacc97}
C.~Caccamo, D.~Costa, and A.~Fucile, {\em J. Chem. Phys.}, \textbf{106}, 255 (1997).
M.~Hasegawa and K.~Ohno, {\em J. Chem. Phys.}, \textbf{111}, 5955 (1999).

\bibitem{n:ruof93}
R.~S. Ruoff, J.~Tersoff, D.~C. Lorents, S.~Subramoney, and B.~Chan,
{\em Nature}, \textbf{364}, 514 (1993).
T.~Hertel, R.~E. Walkup, and P.~Avouris,
{\em Phys. Rev. B}, \textbf{58}, 13870 (1998).

\bibitem{n:rapo97}
L.~Rapoport, Y.~Bilik, Y.~Feldman, M.~Homyonfer, S.~R. Cohen, and R.~Tenne,
{\em Nature}, \textbf{387}, 791 (1997).
M.~Chhowalla and G.~A.~J. Amaratunga, {\em Nature}, \textbf{407}, 164 (2000).

\bibitem{n:gola99}
Y.~Golan, C.~Drummond, M.~Homyonfer, Y.~Feldman, R.~Tenne, and
  J.~Israelachvili, {\em Adv. Mat.}, \textbf{11}, 934 (1999).
Y.~Golan, C.~Drummond, J.~Israelachvili, and R.~Tenne,
{\em Wear}, \textbf{245}, 190 (2000).

\bibitem{uss:schw00d}
U.~S. Schwarz, S.~Komura, and S.~A. Safran, 
{\em Europhys. Lett.}, \textbf{50}, 762 (2000).

\bibitem{uss:schw00e}
U.~S. Schwarz and S.~A. Safran, {\em Phys. Rev. E}, \textbf{62}, 6957 (2000).

\bibitem{n:robe92}
D.~H. Robertson, D.~W. Brenner, and J.~W. Mintmire,
{\em Phys. Rev. B}, \textbf{45}, 12592 (1992).
J.~P. Lu, {\em Phys. Rev. Lett.}, \textbf{79}, 1297 (1997).

\bibitem{n:hern98}
E.~Hernandez, C.~Goze, P.~Bernier, and A.~Rubio,
{\em Phys. Rev. Lett.}, \textbf{80}, 4502 (1998).
G.~Seifert, H.~Terrones, M.~Terrones, G.~Jungnickel, and T.~Frauenheim,
{\em Phys. Rev. Lett.}, \textbf{85}, 146 (2000).

\bibitem{b:land70}
L.~D. Landau and E.~M. Lifshitz,
{\em Theory of elasticity} (Pergamon Press, 1970).

\bibitem{n:srol94}
D.~J. {Srolovitz}, S.~A. {Safran}, and R.~{Tenne},
{\em Phys. Rev. E}, \textbf{49}, 5260 (1994).

\bibitem{n:ters92}
J.~{Tersoff}, {\em Phys. Rev. B}, \textbf{46}, 15546 (1992).
T.~A. {Witten} and H.~{Li}, {\em Europhys. Lett.}, \textbf{23}, 51 (1993).

\bibitem{n:gast83}
A.~P. Gast, C.~K. Hall, and W.~B. Russel,
{\em J. Coll. Inter. Sci.}, \textbf{96}, 251 (1983).

\bibitem{n:teje94}
C.~F. Tejero, A.~Daanoun, H.~N.~W. Lekkerkerker, and M.~Baus,
{\em Phys. Rev. Lett.}, \textbf{73}, 752 (1994) and
{\em Phys. Rev. E}, \textbf{51}, 558 (1995).

\bibitem{comment}
This estimate is valid for weak adhesion. In \cite{uss:schw00d}, we
used an estimate for strong adhesion, $E_{vdW} \sim (A / D^2) R H$.
Then the predicted deformation is larger by a factor of $R/D$, yet the
general conclusions are the same.

\end{thebibliography}
\end{document}